# Pansori: ASR Corpus Generation from Open Online Video Contents

Yoona Choi
Yongsan International School of Seoul
Seoul, South Korea
yoona@ieee.org

Bowon Lee
Department of Electrical Engineering
Inha University
Incheon, South Korea
bowon.lee@inha.ac.kr

*Abstract*—This paper introduces Pansori, a program used to create ASR (automatic speech recognition) corpora from online video contents. It utilizes a cloud-based speech API to easily create a corpus in different languages. Using this program, we semi-automatically generated the Pansori-TEDxKR dataset from Korean TED conference talks with community-transcribed subtitles. It is the first high-quality corpus for the Korean language freely available for independent research. Pansori is released as an open-source software and the generated corpus is released under a permissive public license for community use and participation.

I. INTRODUCTION

Speech has become one of the primary interfaces to access information and use online services through mobile devices and smart home appliances. The conversational interface it provides can enable more natural and faster access than monitor and keyboard.

The accuracy of ASR (automatic speech recognition) has been greatly improved due to the advances in machine learning algorithms [1, 2, 3] in conjunction with the continued efforts to create high quality speech corpus to train ASR models [4, 5].

ASR corpora for English and other Western languages have been developed and made available for independent research over the last few decades [4, 5, 6, 7, 8]. However, it is difficult to find high quality datasets available for open access by academic and technical community.

This paper presents Pansori, a new software tool to systematically create an ASR corpus from online video contents, and Pansori-TEDxKR, a high-quality ASR corpus in Korean, which is based on TED and TEDx [9] conference talks with Korean speech and subtitle data generated and validated by community volunteers.

Pansori increases the quality of corpus generation by utilizing subtitle timing information with alignment adjust and by validating audio-text matches with state-of-the-art speech recognition technology of Google Cloud Speech-to-Text API [10]. Pansori is released as an open source software[1] under MIT license [11] and the Pansori-TEDxKR corpus it generated is freely available[2] for research and development under the same license as the original TEDx contents, i.e. CC (Creative Commons) BY-NC-ND 4.0 license [12].

The contributions of this paper are as follows: (1) it presents Pansori, an easy-to-use tool to generate ASR corpus from online video contents, released as open source for the first time to the best of our knowledge; (2) Pansori is also the first tool in the literature to utilize a cloud-based speech API for the simplified generation of ASR corpus in different languages (~120 different languages and language variants with Google Cloud Speech-to-Text API); and (3) it presents Pansori-TEDxKR, a Korean language ASR corpus generated by Pansori, as the first such dataset released and made freely available under a permissive public access license in Korea to the best of our knowledge.

Section II presents the background of our work. Section III describes our approach to ASR corpus generation in detail. In Section IV, we present the result of ASR corpus generation with open online video contents from TEDx conference talks in Korean as the first case study. Section V concludes the paper with our plan for future work.

II. BACKGROUND

The development of common ASR corpora has played a pivotal role in the development of speech technology. Many efforts have been made beginning as early as in 1993 from the TIMIT (Texas Instrument / Massachusetts Institute of Technology) database [6]. Large scale corpora like the WSJ (Wall Street Journal) [7] and Switchboard [8] databases have enabled the development of LVCSR (large vocabulary continuous speech recognition system) in the domain of broadcast news and telephone conversations, respectively. Research institutes and companies have actively used these corpora in conjunction with privately developed in-house datasets in order to train ASR models for improved accuracy and validating results with a predefined performance metric.

---

[1] https://www.github.com/yc9701/pansori
[2] https://www.github.com/yc9701/pansori-tedxkr-corpus

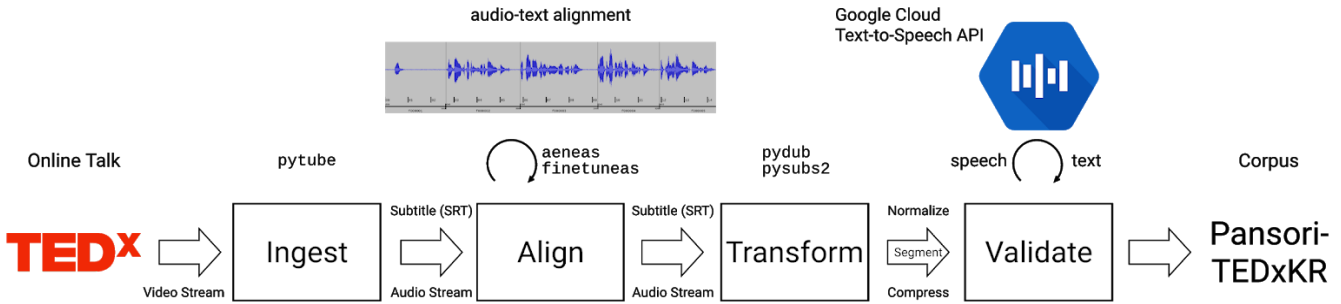

Fig. 1. Processing Steps for ASR Corpus Generation in Pansori.

However, one issue with these datasets is that most of them have been managed by commercial entities like the LDC (Linguistic Data Consortium) [13] and ELRA (European Language Resource Association) [14] consortia and can only be obtained through membership or individual purchase of license for a fee. While this can be considered reasonable for someone who has already started R&D in speech technology in institutional settings, the high cost of collecting, maintaining and distributing high quality corpus data could pose a high entrance barrier to individuals who want to begin independent research and development in speech technology.

In recent years, this dynamic has changed. Several open-source corpora have been created, such as Librispeech, which draws on open-source audiobook data, and TED-LIUM, which takes data from TED talks. These datasets were subsequently released free of charge for developers to use in creating and improving speech recognition technologies. The availability of data now allowed for development to move much more quickly in available languages.

Although development of ASR technology has been progressing in Western languages, the same cannot be said for all languages in the world due to technical, economic, and cultural reasons. Korean is one example of a language for which data resources freely available for ASR research still remains scarce. The challenge is to now create a systematic way to create open corpora in a way that is both cost effective and results in high quality data.

### III. OUR APPROACH IN PANSORI

As described in Fig. 1, our approach to creating ASR corpus from online video contents consists of four key steps: (1) ingest; (2) align; (3) transform; and (4) validate. Each step is implemented as a separate pipeline stage with simple python codes and scripts.

#### A. Ingest

Open online video contents like TED conference talks consists of multiple media streams for different screen resolutions and audio-only playback. Subtitle information hand-transcribed by community volunteers can also be retrieved if available. Pansori retrieves two streams, audio and subtitle data in the SubRip format [15] from online video sharing service via APIs. Pansori uses a Python library called pytube [16] for downloading both audio and subtitle streams.

The downloaded audio and subtitle streams are converted to appropriate formats (audio: mp4 → mp3; subtitle: srt → json) and stored as two separate files for later processing stages.

#### B. Align

The subtitle data contains segmented text corresponding to the audio-visual contents of the associated online video. The following shows two sample segments of a typical subtitle file.

```
1
00:00:15,761 --> 00:00:17,129
오늘 제가 얘기할 주제는요

2
00:00:17,129 --> 00:00:20,337
예술가가 되자. 지금 당장! 입니다.
```

Each segment of subtitle starts with the segment number, followed by timing information and then actual subtitle text.

With the timing information contained in the subtitle file, it is possible to segment the audio stream accordingly in order to make a matching pair of audio and text fragments to be used as ASR corpus. While the timing information provides a very useful feature with which to segment audio streams, inaccuracies can be introduced because the timing information in subtitle data might be determined not only by audio contents, but also by scene changes in the video. In addition, inaccuracies can also arise due to unintentional slicing of audio stream at word boundaries in fast speeches and when substantial ambient noise such as applause is present.

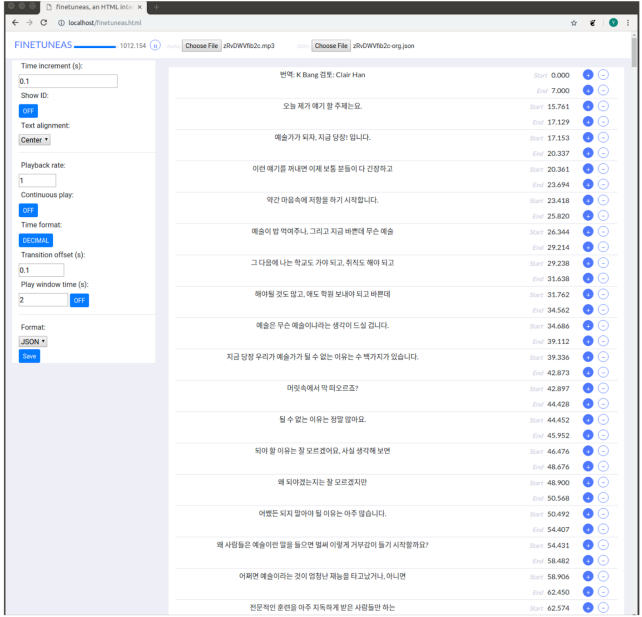

Fig. 2. Speech-Subtitle Alignment System (Finetuneas).

In Pansori, we used the aeneas [17] and finetuneas [18] tools to perform speech-text alignment. Although we could not make the automatic forced alignment feature of aeneas work for Korean language due to the unavailability of good quality Korean TTS software, we found that the finetuneas tool is very useful in reducing the amount of human efforts in fine tuning the speech-text alignment with its intuitive user interface. Fig. 2 shows a sample screen for improving alignment between speech and subtitle data using this tool.

*C. Transform*

The audio stream and subtitle data aligned with each other are then processed with the following transformations specific to data types:

- Audio stream: segmentation, lossless compression
- Subtitle data: normalization, punctuation removal, removal of non-speech text (such as the description of audience response or ambient noise)

Pansori uses Python libraries called pydub [19] and pysubs2 [20] for the segmentation of audio streams according to the timing information contained in the subtitle data.

*D. Validate*

Although the audio stream and subtitle data is force-aligned with each other in the *"Align"* step to make individual audio segments match better with the text labels, there are also inherent discrepancies between the two. This can be caused by one or more combinations of the following: inaccuracies in transcription, ambiguity in pronunciation, and non-ideal audio conditions like ambient noise or poor recording quality.

It is important to refine the corpus by filtering out inaccurate audio-text pairs in the candidate dataset in order to increase the quality of ASR models that will be trained and validated with this corpus.

The previous approaches relied on custom ASR models to validate and refine the generated corpus [4, 5]. Custom ASR models, however, cannot be easily created for many different languages, especially for the languages which do not have existing base of freely accessible ASR corpora.

In Pansori, we took a new approach to use cloud-based ASR service in order to refine the corpus by filtering out audio-text pairs that has high likelihood of being inaccurate. We chose to use Speech-to-Text API of Google Cloud for this [10] because it provides highest quality ASR services in as many as 120 different languages.

The use of cloud service in ASR corpus generation also makes the development of corpus generation system much faster and the configuration and deployment much easier, because it only takes setting up API keys of cloud service instead of setting up custom ASR engines with acoustic models and language models in different languages.

IV. PANSORI-TEDxKR CORPUS

We chose TEDx conference talks [9] as the first source for the ASR corpus we generate using Pansori. TED talks are ideal candidates for this because they are presented in clear speech and contains contents in various topics in Technology, Education, Design as its name stands for. Although majority of TED talks are presented in English, community organized TEDx events are usually presented in the local language of the country.

For Korean language, we identified 41 TEDx talks which contain annotated subtitle data transcribed by TED translators. Table I (Appendix) provides summary of these TEDx talks. The average length of the individual talks is 17 minutes 28 seconds, and the total length of the talks is over 11 hours 56 minutes. Some level of cultural and linguistic diversity can be considered present in the data not only because of their diverse topics but also because of the event locations distributed over different regions of South Korea: Seoul (14), Busan (14) and Daejeon / Daedeok (13). However, talks presented in local dialects of more regions would be desirable for linguistic diversity in the future. As shown in the table, the current dataset is not equally balanced in terms of the diversity of speakers' gender[3].

Using Pansori, we could generate high quality ASR corpus from this TEDx conference talk data. Out of the total 11,704 fragments of the 41 TEDx talks, Pansori identified 3,091 fragments to be included in the Pansori-TEDxKR corpus, based on the quality of the matching of their audio and text. The resulting ASR corpus is close to 3 hours (2 hours 48 minutes) in audio length (corresponds to 26.4% and 23.6% of the total number of fragments and audio length, respectively).

The size of Pansori-TEDxKR can be considered not sufficient for the development of high quality ASR models, when compared to widely used ASR corpus in English (Librispeech: 1,000 hours, TED-LIUM 3: 452 hours). The main goal of creating and releasing the Pansori-TEDxKR corpus is to validate our approach to building open ASR corpus from freely available contents in Korean, and to provide a starting point for community efforts for open ASR corpus. Pansori-TEDxKR is the first Korean language ASR corpus freely available for independent research to the best of our knowledge.

## V. CONCLUSION

In this paper, we have presented a program which can create speech corpora in different languages. This program was used in the generation of the first open corpus for the Korean language freely available for independent research. The corpus was built using Korean TED talks with community- transcribed subtitles and was improved through forced-alignment and further refinement of audio-text pairs using a cloud-based ASR service.

We plan to increase the accuracy of forced-alignment for Korean language in the future. This will make it possible to eliminate manual adjust for better audio-text alignment. We also aim to expand our work with respect to the scope and length of the generated ASR corpus by using various sources of open online video contents.

---

[3] We plan to address the gender diversity issue in the dataset as we increase the size of the ASR corpus with additional data.

[Appendix] Table I. Summary of source video contents for Pansori-TEDxKR.

| Title | Speaker | Gender | Year | Location | Source Video | | Generated Corpus | | Yield (Corpus / Source Data) |
|---|---|---|---|---|---|---|---|---|---|
| | | | | | # Segments | Duration | # Segments | Duration | |
| Appropriate technology | 이성범 | M | 2010 | Seoul | 142 | 10:25 | 87 | 5:58 | 57.8% |
| Making a village worth living in | 김혜정 | F | 2012 | Busan | 366 | 19:07 | 191 | 9:14 | 48.5% |
| The true owner of land | 남기업 | M | 2012 | Busan | 348 | 17:37 | 155 | 6:43 | 38.2% |
| Starting from where I am | 황두진 | M | 2010 | Seoul | 295 | 17:50 | 117 | 6:41 | 37.6% |
| Telling the new story in the old form | 이자람 | F | 2010 | Seoul | 209 | 21:24 | 92 | 7:50 | 36.6% |
| Dreaming a way to future aerial vehicle from unmanned aircraft | 구삼옥 | M | 2011 | Daedeok | 319 | 21:32 | 121 | 7:34 | 35.3% |
| Misconception about evaluations | 유정식 | M | 2012 | Busan | 413 | 19:28 | 158 | 6:43 | 34.7% |
| Be an artist, right now! | 김영하 | M | 2013 | Seoul | 368 | 16:57 | 131 | 5:47 | 34.3% |
| Communication is recovery | 박임순 | F | 2012 | Busan | 438 | 19:17 | 161 | 6:24 | 33.5% |
| Jeju Olleh | 서명숙 | F | 2010 | Seoul | 379 | 28:07 | 135 | 9:16 | 33.1% |
| DIY OOOSSSZZZ band | 유상준 | M | 2010 | Seoul | 123 | 8:08 | 44 | 2:22 | 29.3% |
| Dynamic biology | 이선희 | F | 2011 | Daedeok | 229 | 17:29 | 68 | 4:44 | 27.2% |
| Active immersion in thinking | 황농문 | M | 2012 | Daejeon | 293 | 18:47 | 84 | 5:01 | 26.9% |
| Becoming a good-earthling | 이현정 | F | 2011 | Busan | 340 | 15:07 | 95 | 3:53 | 25.8% |
| More humane medical experience | 김승범, 정혜진 | M, F | 2010 | Seoul | 299 | 18:22 | 80 | 4:36 | 25.1% |
| Finding new energy to overcome resource limits | 이경수 | M | 2010 | Daejeon | 202 | 18:48 | 53 | 4:43 | 25.1% |
| Which do you love, pictures or camera? | 박희진 | M | 2014 | Busan | 140 | 11:07 | 38 | 2:42 | 24.3% |
| Every citizen is a journalist | 오연호 | M | 2010 | Seoul | 254 | 17:16 | 61 | 4:10 | 24.2% |
| Take time to imagine the world to rights | 윤한결 | M | 2013 | Busan | 446 | 21:07 | 126 | 5:01 | 23.8% |
| With feeling the aesthetics of slowness | 이상은 | F | 2011 | Daejeon | 108 | 17:05 | 29 | 3:45 | 22.1% |
| Beating disabilities to pioneer grassroots journalism | 조주현 | M | 2010 | Daejeon | 159 | 18:19 | 37 | 3:56 | 21.7% |
| Statistics 3.0 | 이인실 | F | 2011 | Busan | 407 | 17:16 | 94 | 3:42 | 21.5% |
| Why Analytical Science? | 정광화 | F | 2011 | Daedeok | 229 | 18:36 | 58 | 3:56 | 21.3% |
| Redefinition of soil and its possibilities | 신근식 | M | 2011 | Busan | 343 | 18:11 | 76 | 3:51 | 21.2% |
| Predict disease with face | 김종열 | M | 2011 | Daedeok | 287 | 20:02 | 72 | 4:08 | 20.8% |
| Sustainable DoReMi | 고건혁 | M | 2010 | Seoul | 382 | 17:31 | 78 | 3:10 | 18.2% |
| ITER, towards the dream of a fusion energy era | 정기정 | M | 2010 | Daedeok | 245 | 19:55 | 45 | 3:35 | 18.1% |
| Winning the world with the 'DID' mindset | 송수용 | M | 2010 | Daejeon | 313 | 19:24 | 66 | 3:19 | 17.2% |
| Social venture is blue ocean | 김정현 | M | 2011 | Busan | 338 | 17:45 | 60 | 2:56 | 16.6% |
| No prerequisite learning, no worry | 신현승 | M | 2012 | Busan | 287 | 18:11 | 49 | 2:44 | 15.1% |
| Passion and challenge | 신창연 | M | 2011 | Busan | 485 | 18:29 | 88 | 2:46 | 14.9% |
| Are science and liberal arts equal? | 김상욱 | M | 2013 | Busan | 421 | 18:24 | 67 | 2:36 | 14.2% |
| Perspective, music and life | 다이나믹듀오 | M | 2012 | Seoul | 291 | 20:28 | 48 | 2:51 | 13.9% |
| 아이티 구호현장에서 발견한 음식의 가치 | 김재학 | M | 2010 | Seoul | 49 | 3:10 | 8 | 0:25 | 13.5% |
| A spirit of sharing information and culture 'CC' | 최진권 | M | 2010 | Daejeon | 76 | 12:54 | 18 | 1:42 | 13.2% |
| Gibbons, long-armed apes | 김산하 | M | 2010 | Seoul | 582 | 20:02 | 73 | 2:22 | 11.8% |
| Never let go of your passion, just keep working on it | 김대식 | M | 2010 | Daejeon | 127 | 17:17 | 23 | 1:50 | 10.7% |
| Inconvenient truth of Korean Web | 김기창 | M | 2012 | Busan | 326 | 17:44 | 37 | 1:52 | 10.6% |
| Statecraft, the art of conducting public affairs | 윤여준 | M | 2010 | Seoul | 373 | 19:49 | 46 | 1:59 | 10.0% |
| Korean traditional hawk hunting | 박용순 | M | 2011 | Daejeon | 209 | 17:33 | 21 | 1:09 | 6.6% |
| Multiple identity diaspora | 김경묵 | M | 2010 | Seoul | 64 | 10:09 | 1 | 0:12 | 2.0% |
| Average | | | | | 285.5 | 0:17:28 | 75.4 | 4:06 | 23.6% |
| Total | | | | | 11,704 | 11:56:11 | 3091 | 2:48:10 | |